# Call Admission Control performance model for Beyond 3G Wireless Networks

Ramesh Babu H.S.[1], Gowrishankar[2], Satyanarayana P.S[3].
Department of Information Science and Engineering, Acharya Institute of Technology[1]
Department of Computer Science and Engineering, B.M.S. College of Engineering,[2]
Department of Electronics and Communication Engineering, B.M.S. College of Engineering,[3]
Bangalore, INDIA

*Abstract—* The Next Generation Wireless Networks (NGWN) will be heterogeneous in nature where the different Radio Access Technologies (RATs) operate together .The mobile terminals operating in this heterogeneous environment will have different QoS requirements to be handled by the system. These QoS requirements are determined by a set of QoS parameters. The radio resource management is one of the key challenges in NGWN.Call admission control is one of the radio resource management technique plays instrumental role in ensure the desired QoS to the users working on different applications which have diversified QoS requirements from the wireless networks . The call blocking probability is one such QoS parameter for the wireless network. For better QoS it is desirable to reduce the call blocking probability. In this customary scenario it is highly desirable to obtain analytic Performance model. In this paper we propose a higher order Markov chain based performance model for call admission control in a heterogeneous wireless network environment. In the proposed algorithm we have considered three classes of traffic having different QoS requirements and we have considered the heterogeneous network environment which includes the RATs that can effectively handle applications like voice calls, Web browsing and file transfer applications which are with varied QoS parameters. The paper presents the call blocking probabilities for all the three types of traffic both for fixed and varied traffic scenario.

1. INTRODUCTION

The recent advances in the wireless networks and mobile devices are inclined towards emerging of ubiquitous computing where the user and application running in the mobile terminal (MT) can enjoy seamless roaming. It is well known that the basic problem in the wireless networks is the scarce of the radio resources. The efficient radio resource management is very essential. The admission control is one of the radio resource management technique this plays dominant role in effectively managing the resources. The admission control in the wireless networks will reduce the call blocking probability in the wireless networks by optimizing the utilization of the available radio resources. The mobile communication environment is featured by moving terminals with different QoS requirements in this current scenario the need of guaranteed QoS. The future users of mobile communication look for always best connected (ABC) anywhere and anytime in the Complementary access technologies like Wireless Local Area Networks (WLAN),Worldwide Inter operability for Microwave Access (Wi-Max) and Universal Mobile Telecommunication Systems (UMTS) and which may coexist with the satellite networks [1- 3].

The mobile communication networks are evolving into adaptable Internet protocol based networks that can handle multimedia applications. When the multimedia data is supported by wireless networks, the networks should meet the quality of service requirements. One of the key challenges to be addressed in this prevailing scenario is the distribution of the available channel capacity among the set of multiple users; those are operating with different bandwidth requirements ensuring the QoS requirements of the traffic.

The existing admission control strategies can handle the resource management in homogeneous wireless networks but are unable to handle the issue in heterogeneous wireless environment. The mobility of the terminals in the mobile communication environment makes the resource allocation a challenging task when the resources are always in scarce. The efficient call admission control policies should be in place which can take care of this contradicting environment to optimize the resource utilization.

The design of call admission control algorithm must take into consideration the packet level QoS parameters like minimum delay, jitter as well as session level QoS parameters like call blocking probability (CBP) and call dropping probability (CDP). The CBP is the probability of denial of accepting the new call and CDP the likelihood of dropping the call by a new access network due to decline of the network resources to an unacceptable level in other words the networks is exhausted with the available resources at which it drops the handover calls. In mobile networks the admission control traffic management mechanism is needed to keep the call blocking probability at a minimal level and another RRM strategy vertical handover plays crucial role in reducing the and call dropping probability in an heterogeneous wireless networks.

The further sections of the paper are organized as follows. The section II discusses on the motivation and related work. Section III focuses on the proposed system model for the call admission control based on multi dimensional Markov chain. The section IV is focused on the traffic model. The simulation results are represented in section V and conclusion and future work is indicated in section VI.

## 2. RELATED WORKS

At present, dissimilar wireless access networks including 2.5G,3G, Bluetooth, WLAN and Wi-MAX coexist in the mobile computing environment, where each of these Radio access technologies offer complementary characteristics and features in terms of its coverage area, data rate, resource utilization and power consumption. With all these there are constant improvements in the existing technologies offering better performance at lesser cost. This is beneficial in both the end users and service provider's perspective.

The idea of benefiting from integrating the different technologies has lead to the concept of beyond International mobile telephony 2000(IMT-2000) wireless networks known as the next generation wireless networks (NGWN). In this heterogeneous environment, the end user is expected to be able to connect to any of the different available access networks. The end user will also be able to roam seamlessly within these access networks through vertical handover mechanisms. The global roaming is supplemented by the existence of IP networks as the backbone which makes the mobile computing environment to grow leaps and bounds and can effectively address the issue with regard to converge limitations is concerned. In this multifaceted wireless radio environment the radio resource management plays major role. The effective utilization of the limited available resources is the challenge. The admission control is one such challenge that a network service provider face to achieve better system utilization and to provide the best QoS to the users of the network.

Call admission control schemes can be divided into two Categories, *local* and *collaborative* schemes [4]. *Local schemes* use local information alone (e.g. local cell load) when taking the admission decision [4, 5, 6]. The *Collaborative* schemes involve more than one cell in the admission process. The cells exchange information about the ongoing sessions and about their capabilities to support these sessions [7, 8].

The fundamental idea behind all collaborative admission control schemes is to consider not only local information but also information from other cells in the network. The local cell, where the new call has been requested, communicates with a set of cells that will participate in the admission process. This set of cells is usually referred to as a cluster. In [9] for example, the cluster is defined as the set of direct neighbours. The main idea is to make the decision of admission control in a decentralized manner. There are good amount of work reported on homogenous wireless networks and single service wireless networks. There are few works in the heterogeneous wireless networks.

The Call admission control in Heterogeneous networks is a real challenge. The varied QoS requirements of multimedia applications and the coexistence of different RATs, facade major challenges in designing CAC algorithms for next generation heterogeneous wireless networks. The challenges are heterogeneous networking, multiple service classes, flexible in bandwidth allocation and cross layer issues based design.

Some of the issues the call admission control mechanism should address and point of interest of our research work are as follows. Firstly, The B3G networks should be able to accommodate the applications and user with different QoS requirements, so the CAC algorithms should be designed handle different classes of service meet the QoS needs of all types of applications. Second, there will be diversity in multimedia applications and mobile users QoS requirements in NGWN, The resource utilization and QoS performance can be improved by adaptive bandwidth allocation. This clearly indicates that the CAC should be designed taking into consideration the flexible bandwidth allocation, where, more resources can be allocated when the there is less traffic and the allocated bandwidth can be revoked when there is congestion.

The NGWN has different RATs coexisting which are with different capabilities and they should cater the varied QoS requirements of multimedia applications admission control with single criteria mat be too trivial, in this prevailing scenario the admission control decision should be based on Multi criteria such that the optimization user satisfaction and selection of optimal RAT is achieved. The multi criteria decision making system is an optimization technique used to analyse the contradicting decision making parameters. The MCDM based decision making systems are generally used in the fields o reliability, financial analysis, social and political related analysis and environmental impact analysis etc.

There are several algorithms proposed on handling the admission control decision making using MCDM in heterogeneous wireless networks. This section discusses one specific admission control algorithm based on multiple criteria on which the further work is planned namely *computation-Intelligence-based CAC*.

The *computation-Intelligence-based CAC* use evolutionary approaches like Genetic Algorithm (GA), fuzzy logic and Artificial Neural Networks(ANN) [10].The Majority of the computational-intelligence-based CAC algorithms incorporate fuzzy logic[11],fuzzy neural[12] and fuzzy MCDM[13-14] methods. There are very few works reported on the usage of Fuzzy Neural Artificial Neural Networks in CAC.

## 3. SYSTEM MODEL

In this paper we propose a novel admission control mechanism for effectively handling the call blocking probability in multi class traffic heterogeneous network environment there by increasing the resource utilization. This would in turn achieve the objective of guaranteeing the user

QoS requirements. The proposed model is able to handle three types of the traffic. The traffic considered for the study involves conversation traffic, interactive traffic and background traffic. The representative applications could be voice calls, Web browsing and file transfer applications respectively. The proposed model is developed keeping in mind the WCDMA, Wi-Fi and Wi-Max

The CAC mechanism proposed is focused only on the system's ability to accommodate newly arriving users in terms of the total channel capacity which is needed for all terminals after the inclusion of the new user. In the case when the channel load with the admission of a new call, was precompiled (or computed online) to be higher than the capacity of the channel the new call is rejected, if not, the new call could be admitted. The decision of admitting or rejecting a new call in the network will be made only based on the capacity needed to accommodate the call.

Here the heterogeneous network which comprises a set of RATs, $R_n$, with co-located cells in which radio resources are jointly managed. Cellular networks such as Wireless LAN and Wi-Max can have the same and fully overlapped coverage, which is technically feasible, and may also save installation cost [4]. H is given as H {RAT 1, RAT 2, RAT k} where K is the total number of RATs in the heterogeneous cellular network. The heterogeneous cellular network supports n-classes of calls, and each RAT in set H is optimized to support certain classes of calls.

The paper presents the analytical model for Call admission control mechanism in heterogeneous wireless networks is modelled using higher order Markov chain as shown in figure2.The study considers that, whenever a new user enters the network will originate the network request at the rate $\lambda_i$ and is assumed to follow a *Poisson process*. The service time of the different class of traffic and types of calls is $\mu_i$. The mean service time of all types of users were assumed to follow negative exponential distribution with the mean rate $1/\mu$. Since Voice traffic is exhibits Erlang distribution, the condition that is considered for simulation is Negative Exponential distribution .The parameters of analytic performance model are also called as Performance model parameters and are *number of virtual channels* (N), *user arrival rate* ($\lambda$), *arrival rate of type1 call* ($\lambda_1$), *arrival rate of type2 call* ($\lambda_2$) *arrival rate of type3 call* ($\lambda_3$) and *service time of the user* ($\mu$).

The total number of virtual channel in the system are N. When the numbers of available channels are below the specified threshold the system will drop the calls. The threshold limit is determined by three positive integers $A_1$, $A_2$ and $A_3$. When the number of available channels falls below the threshold $A_3$ the proposed system will accept only the voice calls and web browsing. When the number of available channels falls below the threshold $A_2$ the proposed system will accept only the voice calls and when the available number of channels falls below the threshold $A_1$, the proposed system will not accept any calls, it reaches the stage where there will be no channels available to allocate to the incoming calls and leads to system blocking. The P (0) is probability that there are no allocated channels in the designated system.

The equations (1) - (3) are lower boundary equations for the system states $P_0$, $P_1$ and $P_2$ respectively

$$\lambda_1 P_0 + \lambda_2 P_0 + \lambda_3 P_0 - \mu_2 P_2 - \mu_1 P_1 - \mu_3 P_3 = 0 \qquad (1)$$

$$\lambda_1 P_1 + \lambda_2 P_1 + \lambda_3 P_1 - \mu_1 P_2 - \mu_2 P_3 - \mu_3 P_3 = 0 \qquad (2)$$

$$\lambda_1 P_2 + \lambda_2 P_2 + \lambda_3 P_2 + \mu_1 P_2 + \mu_2 P_2 - \mu_1 P_3 - \mu_2 P_4 - \mu_3 P_5 = 0 \qquad (3)$$

The equations (4) - (6) are upper boundary equations for the system states $P_n$, $P_{n-1}$ and $P_{n-2}$. They are expressed as

$$P_{n-3}(\lambda_1+\lambda_2+\lambda_3+\mu_1+\mu_2+\mu_3)-\lambda_1 P_{n-4}-\lambda_2 P_{n-5}-\lambda_3 P_{n-6}-\mu_1 P_{n-2}-\mu_2 P_{n-1}-\mu_3 P_n=0 \qquad (4)$$

$$P_{n-2}(\lambda_1+\lambda_2+\mu_1+\mu_2+\mu_3)-\lambda_1 P_{n-3}-\lambda_2 P_{n-4}-\lambda_3 P_{n-5}-\mu_1 P_{n-1}-\mu_2 P_n=0 \qquad (5)$$

$$P_{n-1}(\lambda_1 + \mu_1 + \mu_2 + \mu_3) - \lambda_1 P_{n-2} - \lambda_2 P_{n-3} - \lambda_3 P_{n-4} - \mu_1 P_n = 0 \qquad (6)$$

The repeated states are those which are in-between these states i.e. between upper and lower boundaries based on figure1. The repeated states of the system are represented in a generic form as shown in (7).

$$P_4 (\lambda_1 + \lambda_2 + \lambda_3 + \mu_1 + \mu_2 + \mu_3) - \lambda_1 P_3 - \lambda_2 P_2 - \lambda_3 P_5 - \mu_1 P_6 - \mu_2 P_6 - \mu_3 P_7 = 0 \qquad (7)$$

The equation that can be presumed as the general equation for call blocking probability for traffic type 1 is represented in (8).

$$P_n = \frac{\lambda_1 P_{n-1} + \lambda_2 P_{n-2} + \lambda_3 P_{n-3}}{(\mu_1 + \mu_2 + \mu_3)} \qquad (8)$$

Assuming and $\lambda_1 = \lambda_2 = \lambda_3 = \lambda$ and $\mu_1 = \mu_2 = \mu_3 = \mu$, the call blocking probability for type 1 traffic could be expressed as

$$P_n = \frac{a}{3}(P_{n-1} + P_{n-2} + P_{n-3}) \qquad (9)$$

Similarly, the call blocking probability for type 2 traffic is

$$P_{n-1} = \frac{a}{3}(P_{n-2} + P_{n-3} + P_{n-4}) \qquad (10)$$

And for Type 3 traffic is

$$P_{n-2} = \frac{a}{3}(P_{n-3} + P_{n-4} + P_{n-5}) \qquad (11)$$

The call blocking probability for the overall system traffic

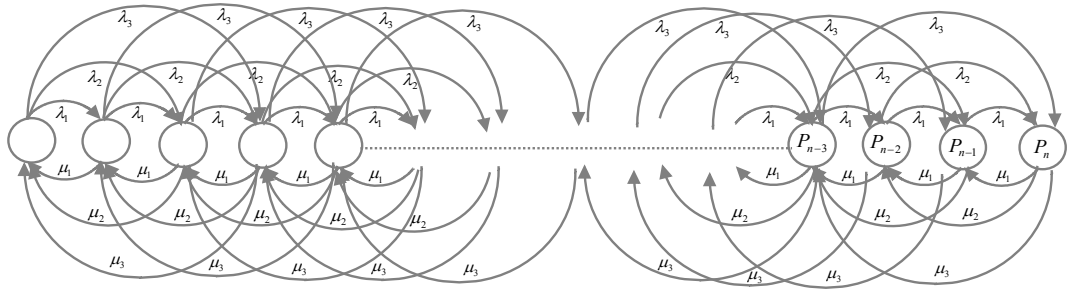

Figure 1: Performance model

The call blocking probability for the overall system traffic $P_{nb}$ can be expressed as

$$P_{nb} = \frac{a}{3}(P_n + P_{n-1} + P_{n-2}) \quad (12)$$

### 4. TRAFFIC MODEL

If we consider a system which is stationary time series system, the series can be modelled as $y(t) = f(x_t)$, where $x_t = $ Where k is the number of factors determines each element of the time series, hence the no stationary time series can be modelled as shown in (13) and (14).

$$y(t) = \sum_{i=1}^{k} p_i^t f_i(t) \quad (13)$$

$$\sum_{i=1}^{k} p_i^t = 1 \quad (14)$$

Where $f_i(t)$ is a random process and hence the non stationary time series can be modelled by a set of random processes[15]. Since the traffic at AP or BS is a non stationary time series, this can be determined by number of users and their arrival pattern, numbers of sessions(application) of individual user, session inter arrival and size of each session. All these are random processes [16].

$$Tr(t) = \sum_{i=1}^{3} p_i^t f_i(t) \quad (15)$$

Where $f_1(t)$ is random process of number of users and the user arrival pattern, $f_1(t)$ is discrete value continuous time process User distribution either uniform or lognormal process and user arrival pattern is time varying Poisson process [16] [17] [18]. The $f_2(t)$ is random process of number of sessions and session inter arrival pattern. The $f_2(t)$ is a discrete value and continuous time random process. User session is Lognormal process or time varying Poisson process and the session inter arrival is a Bi Pareto, Wei bull, Markov Modulated Poisson Process (MMPP) or Time varying Poisson process [15] [17]. The $f_3(t)$ is the size of each session, $f_3(t)$ is discrete value continuous time process and either Bi Pareto or Lognormal random Process [15]. *[y (t-1),y(t-2),…y(t-m)]*. However this is not true in case of non stationary time series, the entire series cannot be determined by single function *f(.)*, instead by set of function $f_1, f_2, ..... f_k$.

### 5. SIMULATION RESULTS AND DISCUSSION

In this section, we present the numerical results and compare the call blocking probabilities of the different types of traffic. The experiment setup is conducted keeping 2 types of traffic constant varying the other type. The second set of experiment setup is conducted varying all the three types of traffic.

The first set of experiments is indicated by the simulation result shown in figure 2. The call blocking probability for a system with N channels which supports three types of traffic is conducted. The experiment considers that, whenever a new user enters the network will originate the network request at the rate $\lambda_1$ for type1 traffic, and $\lambda_2$ for type2 traffic and $\lambda_3$ for type3 traffic and is assumed to follow a Poisson process. The service time of the different types of traffic based calls is considered as $\mu_1$ for type1 traffic, $\mu_2$ for type3 traffic and $\mu_3$ for type3 traffic and is assumed to follow a Lognormal random Process. For the first set of experiments we have considered the arrival rate of all the three types of traffic as $\lambda$ and service rate of all the three type of calls is same and is equal to $\mu$.

The arrival rate of the calls was taken as the varying traffic intensity of Type1 traffic and blocking probability of the type 1, type2, type3 traffic, and overall call blocking probability of the system is plotted. The Figure 2 shows call blocking probability for all three types of traffic when then intensity of the type1 traffic is increased. The horizontal axis shows the number of users with type1 traffic while the vertical axis shows the call blocking probability of all types of traffic.

The simulation results show that the call blocking probability of the system and different types of traffic will increase with the increase in the intensity of type1 traffic. The simulation results with increase the intensity of type2 traffic and simulation results with increase the intensity of type3 also showed the similar kind of results.

The second set of experiments conducted will present the numerical results and compare the call blocking probabilities of the different types of traffic. The proposed a performance model for call admission control mechanism in the heterogeneous RATs and analysing the call blocking probability keeping the variation in the number of channels was conducted . The experiment setup is conducted considering the varying the traffic intensity of type1 traffic the blocking probability of type 1 and blocking probability of the type 2, and type3 traffic of the system is plotted. The Figure 3 shows call blocking probability for all three types of traffic when then intensity of the type 1 traffic is increased. The horizontal axis shows the number of users with type 1 traffic while the vertical axis shows the call blocking probability of all types of traffic.

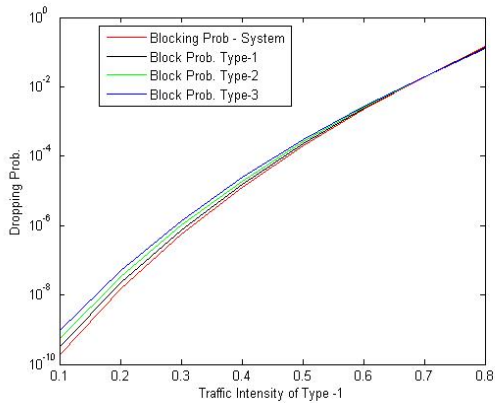

Figure 2. Call blocking probability of the system

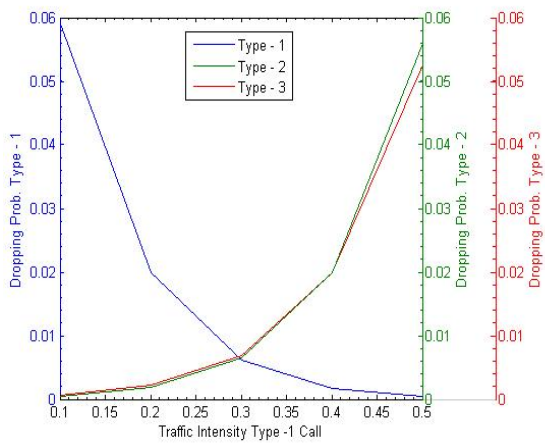

Figure 3. call blocking probablity of varying traffic

The parameters of analytic performance model are called as Performance model parameters are number of virtual channels ($N$), user arrival rate ($\lambda$), arrival rate of type 1 call ($\lambda 1$), arrival rate of type 2 call ($\lambda 2$.) arrival rate of type 3 call ($\lambda 3$) and service time of the calls is taken as $\mu_1$, $\mu_2$ and $\mu_3$.

The simulation results show that the call blocking probability of the different types of traffic will increase with the increase in the intensity of type1 traffic. The simulation results with increase the intensity of type2 traffic and simulation results with increase the intensity of type3 also showed the similar kind of results. The simulation results indicate that at particular state the call blocking probability of all three types of traffic will be minimal.

## 6. CONCLUSION AND FUTURE WORK

In this paper, we have proposed a performance model for call admission control mechanism in the heterogeneous RATs and analysing the call blocking probability keeping the variation in the number of channels.

In order to measure the call blocking probability of the analytical model the simulation study was conducted and following observations were recorded. Firstly, increase in the number of type 1 users will increase the call blocking probability of type2 and type 3 calls and vice versa. Second, Increase in the traffic intensity of one type of traffic will increase the system blocking probability.

The concept of minimizing the call blocking probability is an optimization technique to provide fair QoS to the set of users in the wireless network and there is also a need of intelligent call admission control strategy in the admission control mechanism to make the decision of accepting are rejecting a call keeping the blocking probability minimal in a heterogeneous RATs based network working under dynamic network condition. The future work of this research includes applying intelligence for the process of decision making .The future works includes the use of Neural Network (NN) based decision making based on different criteria in making the decision of admitting or rejecting the call.

## AUTHORS

**Ramesh Babu.H.S** is with the Information Science and Engineering Department, Acharya Institute of Technology, Visvesvaraya Technological University, Soladevanahalli, Bangalore-560 090, Karnataka, INDIA (e-mail: rameshbabu@acaharya.ac.in)

**Dr.Gowrishankar** is with the Computer Science and Engineering Department, B.M.S. College of Engineering, Visvesvaraya Technological University, P.O. Box. 1908, Bull Temple Road, Bangalore-560 019, Karnataka, INDIA (e-mail: gowrishankar.cse@bmsce.ac.in)

**Dr.P.S.Satyanarayana** was with the Electronics and communication Engineering Department, B.M.S. College of Engineering, Visvesvaraya Technological University, P.O. Box. 1908, Bull Temple Road, Bangalore-560 019, Karnataka, INDIA (e-mail: pss.ece@bmsce.ac.in).